\documentclass[journal=jacsat,manuscript=article]{achemso}
\usepackage[english]{babel}
\PassOptionsToPackage{dvipsnames}{xcolor}
\usepackage{chemformula} 
\usepackage[T1]{fontenc} 
\usepackage{booktabs}
\usepackage{subfig}
\usepackage{enumitem}
\usepackage{lineno}
\usepackage{graphicx,psfrag,amsmath,amssymb,amsfonts,latexsym,color,dcolumn,bm}
\usepackage[normalem]{ulem}
\usepackage[hidelinks]{hyperref}
\SectionNumbersOn
\usepackage{soul}
\usepackage{tikz}
\usetikzlibrary{calc}
\usetikzlibrary{decorations.pathmorphing}
\newcommand{\mono}[1]{\texttt{#1}}

\title{Generative Chemical Language Models for Energetic Materials Discovery}
\author{Andrew Salij}
\email{asalij@lanl.gov}
\affiliation{Theoretical Division, Los Alamos National Laboratory, Los Alamos, New Mexico 87545, USA}
\author{R. Seaton Ullberg}
\affiliation{Theoretical Division, Los Alamos National Laboratory, Los Alamos, New Mexico 87545, USA}
\author{Megan C. Davis}
\affiliation{Theoretical Division, Los Alamos National Laboratory, Los Alamos, New Mexico 87545, USA}
\author{Marc J. Cawkwell}
\affiliation{Theoretical Division, Los Alamos National Laboratory, Los Alamos, New Mexico 87545, USA}
\author{Christopher J. Snyder}
\affiliation{Weapon Stockpile Modernization Division, Los Alamos National Laboratory,  Los Alamos, New Mexico 87545, USA}
\author{Cristina Garcia Cardona}
\affiliation{Computing and Artificial Intelligence Division, Los Alamos National Laboratory,  Los Alamos, New Mexico 87545, USA}
\author{Ivana Matanovic}
\affiliation{Theoretical Division, Los Alamos National Laboratory, Los Alamos, New Mexico 87545, USA}
\author{Wilton J. M. Kort-Kamp}
\affiliation{Theoretical Division, Los Alamos National Laboratory, Los Alamos, New Mexico 87545, USA}
\email{kortkamp@lanl.gov}
\setkeys{acs}{keywords = true}
\begin{document}

\maketitle
\pagebreak
\begin{abstract}
The discovery of new energetic materials remains a pressing challenge hindered by limited availability of high-quality data. To address this, we have developed generative molecular language models that have been pretrained on extensive chemical data and then fine-tuned with curated energetic materials datasets. This transfer-learning strategy extends the chemical language model capabilities beyond the pharmacological space in which they have been predominantly developed, offering a framework applicable to other data-spare discovery problems. Furthermore, we  discuss the benefits of fragment-based molecular encodings for chemical language models, in particular in constructing synthetically accessible structures. Together, these advances provide a foundation for accelerating the design of next-generation energetic materials with demanding performance requirements.
\end{abstract}

\pagebreak
\section{Introduction}

Civilian and military applications frequently require the storage and release of chemical energy in energetic materials (EMs)~\cite{badgujar2008advances, zlotin2021advanced}, driving the discovery of successive generations of EM compounds, from the widespread TNT (trinitrotoluene) and RDX (hexogen) in the first half of the 20th century to more recent examples such as TNAZ (1,3,3-trinitroazetidine) and FOX-7 (1,1-diamino-2,2-dinitroethylene)~\cite{redner2010history}. To be effective in operational conditions, these materials must balance multiple properties such as high detonation velocities, thermal stabilities, and a desired insensitivity to shock waves, posing a persistent challenge for EM design.

To avoid the substantial time, financial, and potential environmental costs of manual approaches to EM discovery, computational methods can efficiently evaluate many potential EM candidates~\cite{politzer2001computational}. Machine learning (ML) techniques have proven effective in connecting molecular structure to properties for EMs~\cite{yuan2021materials,zang2022prediction}, highlighting the promise of artificial intelligence for materials discovery. Generative deep learning models have also begun to be explored for this domain~\cite{li2022correlated}, though progress has been constrained by the scarcity of high-quality EM datasets~\cite{zang2022prediction}. Machine learning models ranging from those built using non-linear regression methods \cite{barnes2018machine,davis2024machine}to multi-layer perceptrons \cite{nefati1996prediction} to direct message-passing neural networks \cite{lansford2022building} have demonstrated success in predicting EM properties such as detonation velocities and impact sensitivity. 

While predictions of properties from structures present opportunities in validating candidate materials, the reverse task of producing materials from an initial set of desired properties, or ``inverse design'' of materials~\cite{wang2022inverse}, enables the determination of materials from desired properties. Thus far, deep learning methods such as generative adversarial networks~\cite{nguyen2022synthesizing} and convolutional neural networks~\cite{nguyen2022physics} have been applied for inverse design of systems relevant for EM deployment such as porous structures \cite{choi2023artificial}. The inverse design generation of new molecules for energetic material systems, however, has remained relatively unaddressed, with notable exceptions in the recurrent neural networks (RNNs) developed by Xuemei Pu and coworkers \cite{li2022correlated} and a variational autoencoder produced by Wespiser and Mathieu \cite{wespiser2023application}. Such approaches contribute toward the discovery of new molecules, but the complexities of molecular structure may require additional ML techniques to be properly expressed. For instance, recent advancements in large language models (LLMs) point toward language-based generative methods that leverage textual molecular encodings and transformer attention blocks \cite{vaswani2017attention} as a more flexible and data-efficient framework for advancing EM design. 

Chemical language models (CLMs) have recently emerged as tools for exploring chemical space in a general and scalable manner~\cite{grisoni2023chemical}. Here, we will view CLMs as specialized variants of general-purpose LLMs, which learn to predict sequences in natural language~\cite{naveed2025comprehensive}. By focusing on molecular strings rather than words, permitting a truncated vocabulary and model size, CLMs hold the potential to achieve equal or greater quality at a fraction of the computational cost of general-purpose LLMs. By translating molecular representations into text sequences such as the Simplified Molecular Input Line Entry System (SMILES~\cite{weininger1988smiles}) or the Self-Referencing Embedded Strings (SELFIES~\cite{krenn2020self}), techniques honed for natural language processing become available for molecules. The three-dimensional data of the positions of atoms are compressed onto a one-dimensional textual representation that preserves bonds and bond order.  Such molecular text inevitably discards some geometric information such as absolute atomic positions or relative orientation of unconnected groups, but it retains the essential chemical connectivity required for molecular modeling.

The design paradigm of the generative pre-trained transformer (GPT)~\cite{radford2018improving}, first developed for natural language generation, has recently been extended to small molecules and \textit{de novo} molecular design~\cite{bagal2021molgpt,wang2023cmolgpt}. In these models, molecular text is broken into subunits, or “tokens,” which are linked through an attention mechanism that enables sequential generation of text and the completion of partial inputs~\cite{vaswani2017attention,radford2018improving}. GPTs for chemistry, and transformer-based CLMs more broadly, provide robust and scalable alternatives to other generative approaches such as variational autoencoders and generative adversarial networks~\cite{xue2019advances,meyers2021novo}, producing candidate molecules with high uniqueness while maintaining chemical validity~\cite{bagal2021molgpt,noutahi2024gotta}. As transformer-based CLMs train efficiently and can be reliably improved by increasing data and model sizes~\cite{frey2023neural}, they offer a natural paradigm for leveraging large molecular datasets to generate plausible new compounds. Since these models learn internal chemical representations within their trained weights and biases, it is critical to ensure that essential chemical information is preserved not only during encoding and training, but also through model inference and decoding.

Multiple strategies have been developed for converting molecular graphs into text suitable for machine learning. Owing to its long-standing use and general effectiveness, SMILES~\cite{weininger1988smiles} has been a widely adopted representation in CLMs~\cite{mswahili2024transformer}. However, perceived limitations of SMILES, such as its fragility to mutation, have motivated alternative encodings, including DeepSMILES~\cite{o2018deepsmiles}, SAFE~\cite{noutahi2024gotta}, and SELFIES~\cite{krenn2020self}. SELFIES enforces valency constraints on decoded graphs to preserve molecular validity following small changes to already  valid structures~\cite{krenn2020self}. For CLMs, SELFIES-based models have demonstrated performance comparable to, and in many cases better than, SMILES-based approaches across a range of tasks~\cite{yuksel2023selformer}, underscoring the advantages of alternative molecular encodings for chemical language modeling.

As molecules often contain recurring motifs such as conjugated rings and functional groups, directly encoding fragments has been proposed to increase the semantic meaning of each unit in textual representations. Approaches such as SAFE~\cite{noutahi2024gotta} and GroupSELFIES~\cite{cheng2023group} introduce vocabularies of chemically meaningful fragments, thereby improving the semantic richness of the encoding step. Much as morphemes or words convey more information than individual letters, fragments provide higher-level units than atoms or bonds, with the potential to improve downstream models. For example, the initial GroupSELFIES report showed that synthetic accessibility, defined in terms of the SA score,~\cite{ertl2009estimation} improved for variational autoencoders compared with SELFIES~\cite{krenn2020self}. While GroupSELFIES have since been incorporated into parts of ML workflows~\cite{kelly2025unified}, to our knowledge they have not yet been demonstrated as the primary encoding method in a transformer-based CLM.

The dominant research thrust in developing GPTs for chemistry has centered on drug discovery and pharmacological properties, \textit{not} on EM discovery and its unique design challenges. For example, MolGPT~\cite{bagal2021molgpt}, a pioneering model in \textit{de novo} molecular generation, was trained in part on the MOSES dataset of druglike compounds~\cite{polykovskiy2020molecular} and evaluated using a quantitative estimate of drug-likeness~\cite{bickerton2012quantifying}. More recent models~\cite{noutahi2024gotta,lee2025genmol} have continued this focus on pharmaceuticals, leaving EM discovery comparatively unexplored. Moreover, the massive chemical datasets commonly used for GPT pretraining, such as ZINC~\cite{irwin2005zinc,tingle2023zinc}, were developed largely to characterize pharmaceutical space, which may limit their utility when directly applied to EMs. Since EM-specific databases remain far smaller in scale~\cite{huang2023database}, methods that can leverage general chemical databases while adapting effectively to EMs are essential. 

\begin{figure}[!t]
\includegraphics[scale=.5]{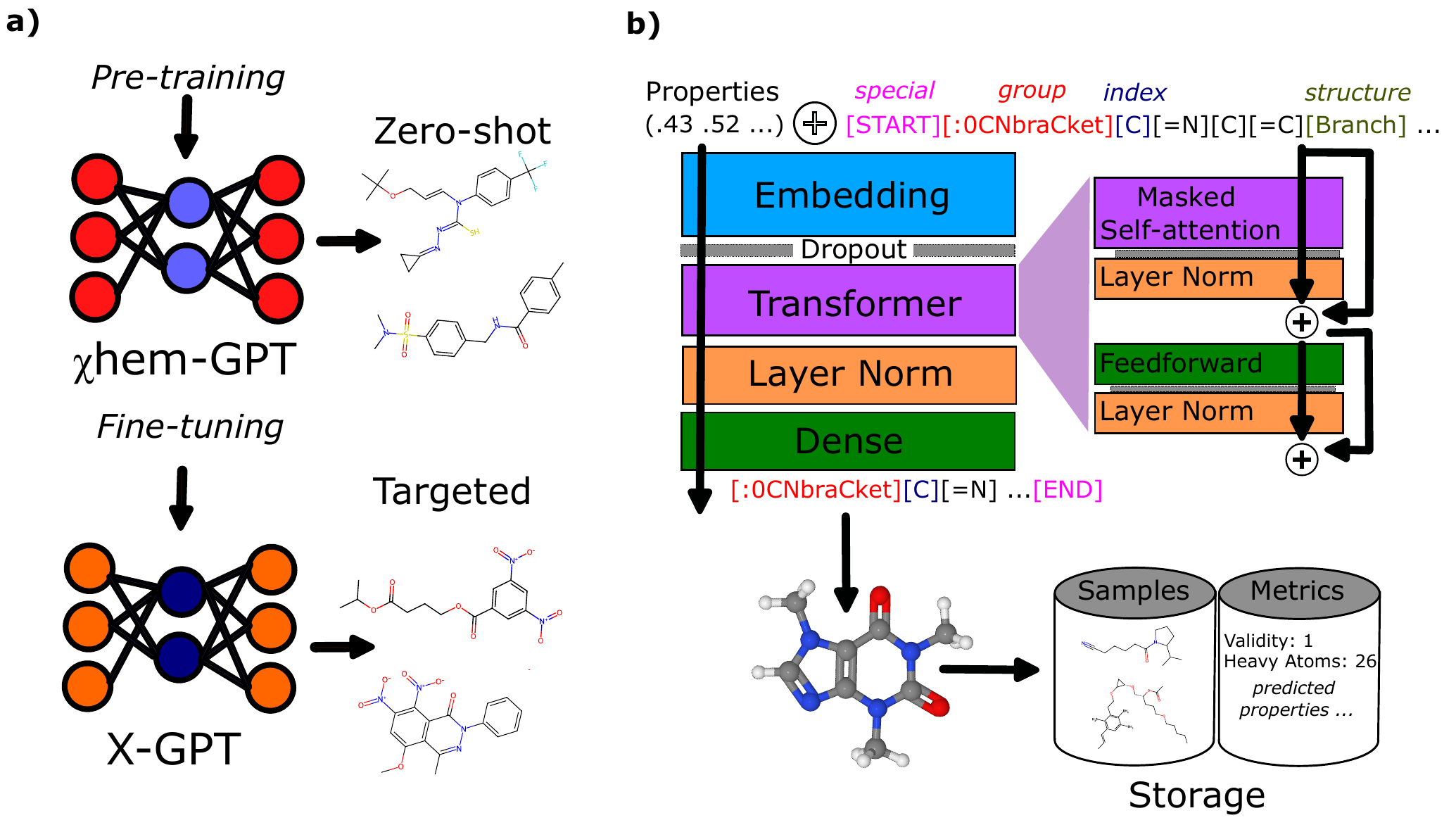}
\caption{a) Training pipeline for GPT models, staged into pretraining to produce a wide variety of molecules and fine-tuning, which produces many C-, N-, O-containing compounds. b) Scheme of GPT model architecture and data processing.}
\label{fig:scheme_overview}
\end{figure}

In this paper, we demonstrate how transfer learning enables a foundational chemistry GPT ($\chi$hem-GPT), pretrained on large molecular datasets to learn a general molecular grammar, to be fine-tuned on a relatively small EM dataset to create X-GPT (Fig.~\ref{fig:scheme_overview}a). This approach directly addresses the minimal sizes of EM datasets by exploiting the scalability of language models, providing a framework for \textit{de novo} generation of EM candidates. Particularly, as EM molecules share structural elements with pharmaceuticals, pretraining here serves to encode relevant molecular vocabulary and grammar before the task-specific focus of fine-tuning. By bridging large, general chemical datasets with small, specialized EM collections, our approach establishes a practical route for generative modeling in domains where data scarcity has historically limited progress. As this generative AI produces generally novel and unique candidates, our transformer models expand the range of traversed chemical space and can augment datasets. Furthermore, we show that transformer-based chemical language models, previously primarily applied to pharmaceutical molecular discovery, have effective application for energetic materials discovery. In doing so, we demonstrate the  efficacy of the adaptation of molecular foundation models to EM inverse design objectives, potentially distributing pre-training computation cost over many post-training design objectives. As such, we have implemented a framework for EM design that is adaptable and scalable, which are engineering objectives that bolster the core scientific goals of molecular discovery.

\section{Methods} 
\subsection{Model Architecture}

Our general chemistry model, $\chi$hem-GPT,  consists of embedding layers for molecular string data as well as potentially feature/property data, a transformer stack, and a linear token output layer (Fig.~\ref{fig:scheme_overview}b). With regards to embedding, we separately encode token indices, absolute positional indices, and sequence indices as integers. Here, ``sequence index'' refers to an indicator as to the type of data input into the model. This distinguishes between input numeric properties and the molecular string data. Numeric properties are concatenated to the left of the inputs, a feature we use for fine-tuning and omit for pretraining. The transformer stack consists of 12 decoder layers with an embedding dimension of $512$ or $1028$ for a ``small'' vs. a ``large'' model and a feed-forward dimension of $4$ times the embedding dimension. In each transformer, a causal multihead self-attention mechanism~\cite{vaswani2017attention} has keys masked according to sequential position and padding and with sequentially later tokens attending to prior ones but not the reverse. Both prior to and within the transformer layer stacks, dropout layers with dropout percentage $p = 1\%$ are present during training. This architecture is broadly reminiscent of the typical ``decoder-only'' transformer architecture~\cite{brown2020language}, though we note that the presence of dropout layers as well as the precise handling of shortcut connections and normalization layers alongside attention layers differs from the canonical implementation~\cite{vaswani2017attention}. Namely, we put the normalization layers \textit{inside} the shortcut connection, an architecture detail that has previously shown improvements for training~\cite{xiong2020layer,olmo20242}.

The models are trained to predict next tokens from a predefined lexicon (see Section \ref{sec:dataset}) over a constant context window $w_{context}$ as they are evaluated via categorical cross-entropy loss. As such, at inference they produce raw logits converted via softmax to a probability distribution of next tokens within the context window from which the next token is multinomially sampled. Prior to the invocation of softmax, the raw model outputs may be divided by a temperature $T$\footnote{Unless otherwise noted, $T = 1$.}, for which higher $T$ samples less-probable tokens more frequently. Here, all inputs are padded to the predefined context window via ``[PAD]'' tokens, which are masked within the attention mechanism but \textit{not} within the loss function (i.e., the model could, in principle, predict a next token as a ``[PAD]'' token).

For our analysis, we will be primarily considering two versions of $\chi$hem-GPT: (i) a ``large'' size model ($\sim150-160$ M parameters, $w_{context} = 512$) with encoding schemes either of a SELFIES or GroupSELFIES variety and (ii)  ``small'' size variants ($\sim 40$ M parameters, $w_{context} = 256$)  in order to explore the effects of scaling (model parameters are summarized in Table~\ref{tab:si_model_parameter}). With regards to SELFIES, we have used the literature implementation \cite{selfies_software} with minimal adjustment. In contrast, we have used the main GroupSELFIES implementation \cite{groupselfies_software} with custom modifications to improve serialization/deserialization of data as well as to enforce an identical indexing scheme to SELFIES. It was reported that the GroupSELFIES  \cite{cheng2023group}  uses a base-16 indexing scheme, but the main implementation \cite{groupselfies_software} uses a base-13 scheme due to the lack of certain tokens in GroupSELFIES that are present in SELFIES (e.g., ``[Branch2]''). To avoid a discrepancy, we have used the default SELFIES indexing alphabet, which contains tokens not native to GroupSELFIES. For the remainder of this paper, we will refer to this modified indexed GroupSELFIES as ``GroupSELFIES $i$'' to distinguish from the base-13 GroupSELFIES. 
\subsection{Dataset}
\label{sec:dataset}

$\chi$hem-GPT is trained on a subset of the SAFE dataset~\cite{noutahi2024gotta,irwin2005zinc,chambers2013unichem} consisting of approximately 8M molecules that converted successfully into SELFIES~\cite{krenn2020self} (see Section~\ref{si:sec_safe_dataset} for . We refer to this collection as ``SAFE-8M’’ throughout this work. The dataset spans a wide range of small-molecule sizes (see Supplementary Materials Fig.~\ref{fig:safe_dist}), providing a broad section of chemical space for the foundation model to learn from. SELFIES vocabularies are constructed by parsing SAFE-8M at the token level. For GroupSELFIES, $1000$ byte-pair encodings~\cite{shibata1999byte} are generated from analogous SMILES strings, and only those that can be independently converted back into valid GroupSELFIES are retained. This procedure yields a grammar linking molecular graphs to single group tokens. To ensure chirality is represented, groups centered on nitrogen, carbon, and phosphorus are manually added, since the byte-pair encoding process does not always preserve stereochemistry. Unlike the approach reported in the original GroupSELFIES paper~\cite{cheng2023group}, our method scales to arbitrary numbers of molecular subgraphs with minimal manual intervention. All models are trained with special tokens, including the start token “[START]’’ and end token “[END],’’ which bracket each full molecule. During unconditioned inference, generation begins with a start token and terminates either when an end token is produced or when a predefined maximum context length is reached. With context windows of 256 or 512, the models reported here generally demonstrate early termination via an end token.

In contrast, X-GPT is fine-tuned on a comparatively small dataset of approximately 17 thousand energetic-like molecules (``X-17K’’) with structures sourced from the Cambridge Structural Database (CSD) ~\cite{allen2002cambridge}.   This dataset was curated by down-selecting all molecules in the CSD that contain at least one carbon atom and have valid three-dimensional structural representations files. Molecules were further filtered to retain only those composed of H, C, N, and O elements with fewer than sixty non-hydrogen atoms. The final X-17K dataset consists of the remaining compounds that contain at least one N–N, N–O, or O–O bond. Estimated detonation properties were determined using the thermochemistry software CHEETAH~\cite{fried1994cheetah} (details on property distributions in Supplementary Materials Fig.~\ref{fig:x_dist}). Necessary inputs, i.e., material density was taken from the CSD database while solid heats of formation were computed using high-throughput density functional theory calculations.  The gas phase heats of formation were first calculated using $\omega$B97X-D/6-311G** level \cite{chai2008long} of theory and then corrected for sublimation enthalpy \cite{mathieu2018accurate} to produce solid heats of formation \cite{SEATON}. Outliers were removed from the database of calculated detonation performance, and the remaining data were linearly renormalized to map properties into the domain $[0,1]$. X-17K represents a curated selection of literature molecules augmented with computed energetic properties, though not all molecules in it exhibit high predicted detonation performance. That is, the distribution of detonation properties for X-17K contain the potential but not the necessity of high performance. Molecules in X-17K differ substantially from those in SAFE-8M: when sampling 1,000 molecules five times from each dataset, we found that their descriptors had a Guacamol KL score~\cite{brown2019guacamol} of $0.630 \pm 0.033$, where a score of $1$ indicates identical distributions and $0$ indicates complete dissimilarity. A summary of both datasets is available in the Supplementary Materials (Table~\ref{tab:si_datasets}) for reference.
\subsection{Training}
\label{sec:training}

The ``small'' $\chi$hem-GPT models were trained on a single GPU node with four NVIDIA A100 Tensor Core GPUs using distributed data parallelism to coordinate data flows. Training was performed for one full epoch on the SAFE-8M dataset, which was split into $80\%$, $10\%$, and $10\%$ for training, validation, and testing, respectively, with a batch size of $b = 250$. The large $\chi$hem-GPT models were trained in a similar manner but on eight GPU nodes (32 A100 GPUs total), using two epochs and a reduced batch size $b = 64$. All models used the Adam optimizer~\cite{kingma2014adam} with an initial learning rate of $0.001$ that decayed linearly to $0$ by the final epoch. Models were trained to minimize categorical cross-entropy loss~\cite{pytorchcrossentropy} on the predicted next-token logits across the entire context window.
%
All models converged within approximately 10 hours under the above specifications, where convergence was defined as a change in validation cross-entropy loss of less than $0.01$ between validation iterations.

The large $\chi$hem-GPT models were fine-tuned on the X-17K dataset for $20$ epochs using two GPU nodes. Alongside the input tokens, the models were conditioned on detonation pressure $P$ and velocity $v$. Fine-tuning used an initial learning rate of $0.0001$ with a linear decay schedule as in pretraining. We explored two strategies: (i) basic fine-tuning, where only the first and last transformer layers and the output layer were unfrozen, and (ii) Low-Rank Adaptation (LoRA)~\cite{hu2022lora,yu2023low}, with rank $r = 8$ and scaling factor $\alpha = 32$. In LoRA, rank-decomposition matrices are introduced as adapters into selected frozen layers (e.g., transformer attention), reducing the number of trainable parameters while preventing the base model from deviating from its initial weights~\cite{hu2022lora}. LoRA has shown promising results in both chemical design~\cite{wang2024pepdora} and general LLMs~\cite{buehler2024x}, though its effectiveness for chemistry remains less established relative to conventional fine-tuning. Applied to our GroupSELFIES-trained large $\chi$hem-GPT, LoRA reduced the number of trainable parameters to $\sim 0.6$M, compared to 156M total parameters in the base model. Unless noted as ``LoRA,'' all fine-tuning results presented are for the ``basic'' fine-tuning.
\subsection{Inference}

As the models are trained to predict the next token in a sequence, prompting with a \textit{start} token produces a distribution of possible continuations within the context window. After applying temperature $T$ and softmax scaling, one option is sampled and appended to the input. The updated sequence is then fed back into the model, which continues generating tokens until either an \textit{end} token is reached or the maximum permitted molecular length is obtained.

%
\subsection{Metrics}

To characterize the generated outputs, we first clean them by parsing special tokens and convert them to SMILES where possible, and then into RDKit \cite{rdkitsoftware} molecule objects. An output is considered ``valid'' if it successfully converts to a RDKit representation with at least one atom. Thus, a failure of SELFIES to map to SMILES results in an ``invalid’’ molecule, even if subsequent post-processing could in principle recover it. This process underlines how the results of generative artificial intelligence systems are products of the entire data pipeline of which neural networks are a critical but not sole component, a fact reflected in our metric definitions.

From the subset of valid molecules, uniqueness is measured as the percentage of distinct canonical SELFIES after data processing. Novelty is assessed by comparing the canonical SMILES of valid molecules against those in the combined training dataset (SAFE-8M and X-17K) for all models. Although the pretrained models were not trained directly on X-17K, the overlapping molecules in that set are relatively small. Defining novelty in this way allows for a consistent comparison with the fine-tuned X-GPT models. For internal diversity, we follow the Molecular sets convention~\cite{polykovskiy2020molecular,benhenda2017chemgan}, namely:
\begin{align}
\text{Diversity}(S) = 1 - \frac{1}{|S|^2}\sum_{x,y\in S} \text{Similarity}(x,y)
\end{align}  
for molecules $x$ and $y$ in a generated set of size $S$, with similarity defined as the Jaccard–Tanimoto coefficient~\cite{jaccard1912distribution,rogers1960computer} between molecular fragments. This definition includes a single count of each molecule with itself, which by definition yields a similarity of 1, leading to a modest downward bias in the overall measure that vanishes as $|S|\rightarrow \infty$. For computational efficiency, statistics are batched and pooled over subsets $s \in S$ of the generated molecules. Fingerprints are computed using RDKit default parameters, which have been largely unchagned for the past decade~\cite{rdkitsoftware2014,rdkitsoftware}. As Tanimoto similarity can be sensitive to particular parameters that are not always specified in the literature, our diversity values may not be perfectly comparable across all prior reports. Nevertheless, the definition here is consistent with those used in MolGPT~\cite{bagal2021molgpt}, GenMol~\cite{lee2025genmol}, and VAE models employing GroupSELFIES encodings~\cite{cheng2023group}. 

Because SELFIES provide a mapping that creates valid molecules with some errors in the text, their validity rate approaches $100\%$~\cite{krenn2020self}, motivating the use of additional metrics to assess the quality of generated molecules. For instance, the synthetic accessibility (SA) score\cite{ertl2009estimation} provides a proxy for how simple a molecule would be for an organic chemist to synthesize where a very high score would be unlikely to be possible to make. Within the SELFIES and GroupSELFIES encodings, we find that the number of radical electrons (denoted ``$N$ Radicals’’) after conversion to SMILES and subsequently to an RDKit molecular graph serves as a useful proxy for molecular quality, with higher radical counts indicating poorer outcomes. These radicals arise as artifacts of SELFIES/GroupSELFIES parsing, which enforces valency, and are not a prominent feature of the training datasets. While such molecules are still considered ``valid'', a low radical count provides a more precise quality criterion. Furthermore, synthetic accessibility, measured using the SA Score~\cite{ertl2009estimation}, offers a general metric for comparison to the training datasets and quality.

An additional goal for CLMs, particularly in unconditioned generation, is to reproduce the underlying distributions of the training data. To evaluate this, we follow the Guacamol benchmarking protocol~\cite{brown2019guacamol}, which defines a Kullback–Leibler divergence over a set of probability distributions $P$ and $Q$ as~\cite{kullback1951information}
\begin{align}
D_{KL} = \sum_i P_i \log \frac{P_i}{Q_i}.
\end{align}
Distributions are computed for $n$ molecular descriptors, and the final score is given by~\cite{brown2019guacamol}
\begin{align}
\text{KL}{\text{score}} = \frac{1}{n}\sum_i^n \exp(-D_{KL}),
\label{eq:kl_score}
\end{align}
where a value of 1 indicates perfect fidelity to the target distribution. Valid molecules are used to construct benchmarks such as uniqueness and novelty, while additional descriptive metrics are provided in the Supplementary Materials (see Section~\ref{sec:si_characterization}).  We trained a surrogate ChemProp model~\cite{heid2023chemprop}, denoted as XChemProp, to rapidly predict detonation properties from SMILES for use in validating GPT outputs (See Supplementary Materials Fig.~\ref{fig:xchemprop}). This surrogate enables efficient benchmarking of molecules generated by our GPT models and provides a useful proxy for development. For final benchmarking we selected representative samples of generated molecules, performed density functional theory calculations on them, and used those results to evaluate detonation velocities using both the CHEETAH software \cite{fried1994cheetah,SEATON} as well as Kamlet-Jacobs equations \cite{kamlet1968chemistry}, which are well-established semi-empirical approaches for calculating detonation properties. For high-throughput characterization of detonation properties, we apply XChemProp to SMILES strings converted from SELFIES outputs from the GPT models as well as performed CHEETAH \cite{fried1994cheetah} and Kamlet-Jacobs \cite{kamlet1968chemistry} directly on a representative subset of generated molecular graphs converted to 3D structures. The surrogate XChemProp enables quick characterization within the training loop, while the latter two methods present results that should prove closer to ground truth in principle.

\section{Results and Discussion}
\label{sec:results}

\subsection{Pre-trained foundational chemical language models}

\begin{table}
\begin{tabular}{llcccccc}
Model (Encoding) & Validity ($\uparrow$) & Uniqueness  ($\uparrow$) & Diversity  ($\uparrow$) & Novelty  ($\uparrow$) \\
\hline
MolGPT \cite{bagal2021molgpt} (SMILES) &$0.994$ & $\mathbf{1.0}$ & $0.857$ & $0.797$\\
GenMol \cite{lee2025genmol} (SMILES) &  $\mathbf{1.000\pm0.000}$ & $0.997\pm 0.001$ &$0.818\pm0.001$& -\\
Group-VAE-125 \cite{cheng2023group} & $\mathbf{1.0(0)}$ &$0.9985(4)$&$\mathbf{0.8587(1)}$ & $0.7187(11)$\\
SELFIES-VAE-125 \cite{cheng2023group}  & $\mathbf{1.0(0)}$ & $0.9986(4)$& $0.8579(1)$& $0.7345(16)$\\
\hline
$\chi$hem-GPT (SELFIES) & $\mathbf{0.999\pm0.001}$ &$\mathbf{1.000\pm0.000}$&$0.820\pm0.001$ &$\mathbf{0.998\pm0.001}$ \\
(LARGE) (GSELFIES $i$) & $0.911\pm0.006$&$0.996 \pm 0.002$ & $0.821\pm 0.001$ &$0.995\pm 0.001$ \\
\hline
$\chi$hem-GPT (SELFIES)& $\mathbf{0.999\pm0.001}$ & $\mathbf{1.000\pm.000}$ & $0.808\pm0.003$& $\mathbf{0.998\pm 0.002}$\\
(SMALL) (GSELFIES $i$)& $0.907\pm0.007$&$0.995\pm.004$&$0.819\pm0.005$& $\mathbf{0.996\pm 0.002}$\\
 \hspace{2cm} (GSELFIES)& $0.999\pm0.000$&$\mathbf{0.999\pm0.001}$&$0.820\pm0.002$ & $\mathbf{0.996\pm 0.001}$\\
\hline 
\end{tabular}

\vspace{10pt}
\begin{tabular}{llcc}
Model & Encoding& SA Score  ($\downarrow$) & $N$ Radicals  ($\downarrow$) \\
\hline
$\chi$hem-GPT & SELFIES &$4.63\pm1.18$ & $0.33\pm0.85$ \\
(LARGE) & GSELFIES & $\mathbf{3.51 \pm 0.99}$&$\mathbf{0.02 \pm 0.16}$\\
\hline
$\chi$hem-GPT & SELFIES &$4.60\pm1.19$&$0.27\pm0.71$\\
(SMALL) & GSELFIES $i$&$3.60\pm1.06$ &$\mathbf{0.02\pm.019}$\\
 & GSELFIES& $3.69\pm0.99$&$0.03\pm0.20$\\
 \hline
\end{tabular}
\caption{Benchmark metrics for the large and small $\chi$hem-GPT models encoded via SELFIES  and GroupSELFIES $i$ compared to prior art in the literature. For $\chi$hem-GPT models, $5000$ molecules have been generated in $N = 5$ isolated runs of $1000$ molecules each. Validity, uniqueness, novelty, and diversity are determined as scalars ranging from $0-1$ for each isolated run, while the other metrics have their means and variances pooled. GroupSELFIES $i$ is GroupSELFIES that has its index alphabet enforced to be identical to SELFIES. Cells for literature models source from their respective papers, and cells are blank where the corresponding metric does not exist.  }
\label{tab:xhem_gpt}
\end{table}

From the initial pretraining, we have produced CLMs that are competitive with previously reported models having a variety of underlying architectures (see Table ~\ref{tab:xhem_gpt}). In particular, we find nearly perfect novelty of $\sim 99 \%$ generated molecules with respect to our training datasets, a substantial improvement over the $\sim70-75 \%$ novelty of SELFIES-based VAEs\cite{cheng2023group} and the $\sim 80\%$ novelty of the pioneering MolGPT \cite{bagal2021molgpt} Namely, for the SELFIES-containing models, we find that uniqueness, validity, and novelty are all nearly $100\%$, though the GroupSELFIES (GSELFIES) models slightly diverge from this. This discrepancy might be due to fixing the GroupSELFIES indexing alphabet to that of the SELFIES one, which can potentially produce invalid GroupSELFIES. As seen in Table \ref{tab:xhem_gpt}, enforcing the base-13 GroupSELFIES indexing ameliorates this, but it modestly increases the synthetic accessibility score. As a lower SA value implies greater accessibility, this suggests that the subset of valid GroupSELFIES $i$ are more accessible than the native GroupSELFIES. Plausibly, the model may fail to distinguish between indexing and non-indexing tokens, but having these all be homographs obscures this. For where indexing tokens do not fully overlap with chemical structure tokens in GroupSELFIES $i$, such token ambiguity can create impossible-to-parse strings that would have been valid for GroupSELFIES. However, such outputs would likely not have been representative molecules compared to the training datasets as they arose from a fundamental error in model syntactical understanding; therefore, excluding them works as a filter for higher quality outputs. Such a description echoes the recent findings that the generation of invalid molecular strings is a desirable characteristic for CLMs \cite{skinnider2024invalid}. 

We find that both pretrained model encodings partially match their training distributions, with KL scores as defined in Eq.\eqref{eq:kl_score} of
$ 0.668 $ for the large $\chi$hem-GPT (SELFIES) and $ 0.847 $ for the GroupSELFIES variant as determined on the entirety of unique, valid examples of $5000$ generated molecules. In this metric, GroupSELFIES proved to be more effective than SELFIES to produce outputs of a given set of qualities. We note that these divergence scores are lower than models in the scientific literature utilizing SMILES \cite{brown2019guacamol,bagal2021molgpt}. We attribute this to SELFIES enforcing constraints on the generated systems that shifts the generated molecules from the target distributions as well as the fact that SAFE-8M is a different dataset than those of the prior reports. Furthermore, despite the large models exhibiting lower loss during pretraining than their small counterparts (see training loss trajectories in Supplementary Materials Section ~\ref{sec:si_training}), they do not perform dramatically better for core quality metrics. While this demonstrates that scaling will not always result in great increases in performance, it highlights the gains to be made by data processing and model design. 
\begin{figure}[b!]
\includegraphics[scale=.9]{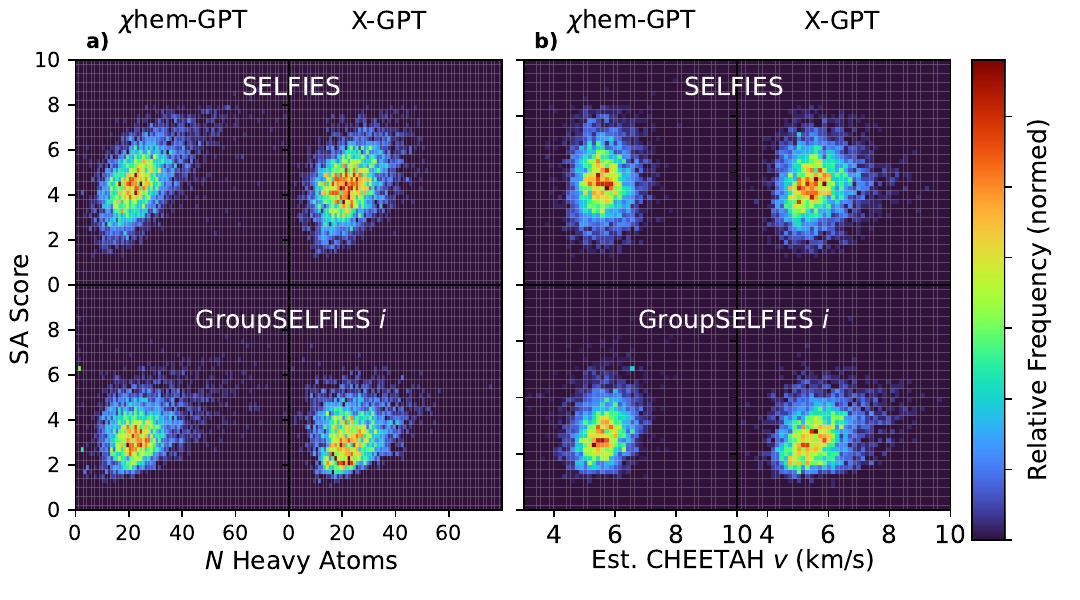}
\caption{a) Synthetic accessibility (SA) score \cite{ertl2009estimation} distributions for unconditioned molecular outputs of pretrained $\chi$hem- and fine-tuned X-GPT models with a) number of heavy atoms generated and b) predicted detonation velocities via ChemProp \cite{heid2023chemprop} surrogate. All subfigures are normalized such that the highest histogram bin is 1.}
\label{fig:sa_comparisons}
\end{figure} 

The GroupSELFIES models produce much more synthetically accessible (i.e., a lower SA score) molecules than the SELFIES based models, dropping from a mean of $4.63$ to $3.51$ for the large $\chi$hem-GPT series (Table \ref{tab:xhem_gpt}). Presumably, the addition of meaningful molecular groups provides the model with a more descriptive vocabulary for which next token prediction works better. As the SELFIES models generate molecules with more heavy atoms, the SA score increases, but this effect is much weaker for GroupSELFIES (Fig.~\ref{fig:sa_comparisons}). For a molecule of a given size, GroupSELFIES requires fewer tokens to predict, making the inference task shorter than for SELFIES and the number of relevant tokens in the context window fewer. While the GroupSELFIES strings do incur additional computational costs in reconversion to molecular graphs, the tradeoff of GPU usage in inference for CPU usage in post-processing proves justified as the total inference and benchmarking time decreases for GroupSELFIES. For instance, generating and benchmarking $1000$ molecules using the SELFIES $\chi$hem-GPT took $139\pm2$ minutes while corresponding GroupSELFIES processes took  $91\pm1$ minutes on similar computational specifications. 
\begin{table}[b!]
\begin{tabular}{lcccccc}
Temp & Validity ($\uparrow$) & Diversity ($\uparrow$) & SA Score ($\downarrow$) & $N$ Radicals ($\downarrow)$& $N$ Heavy Atoms\\
\hline
$0.8$& $\mathbf{0.999\pm0.001}$&$0.811\pm0.003$ &$ \mathbf{4.27\pm 1.16}$& $\mathbf{0.13 \pm 0.47}$& $25.04\pm11.79$\\
$1.0$ & $\mathbf{0.999\pm0.001}$ & $0.820\pm0.001$&$4.63\pm1.18$ & $0.33\pm0.85$ &$23.57\pm9.03$ \\
$1.2$  &$0.994\pm0.003$ & $0.839\pm0.003$  & $4.97\pm1.18$ & $0.67 \pm1.21$&$22.35\pm9.37$\\
$2.0$ & $0.892\pm0.005$ & $\mathbf{0.943\pm 0.002}$ & $6.56\pm 1.19$ &$3.27\pm3.45 $& $13.84\pm 9.40$\\
\hline 
\end{tabular}
\caption{Temperature screening for the large $\chi$hem-GPT model trained using SELFIES over benchmark metrics. Runs and statistics performed equivalently to in Table \ref{tab:xhem_gpt}.\label{tab:xhem_gpt_temp}}
\end{table}

Controlling the temperature applied to the final softmax function used at inference time enables control of the diversity and quality of output molecules as shown in Table \ref{tab:xhem_gpt_temp}. The implementation of SELFIES fails to convert very low quality text such as the improbable strings that manifest at higher temperatures. One such invalid generated SELFIES was ``[O][\textbackslash\textbackslash N-1][Branch1][\# C][/Ge]...[N][As][=As-1]", which runs into excessive valency around the Ge atom. Here, we consider this decrease in validity a useful metric as strings that break the fairly robust SELFIES specification should be considered low quality. Likewise, with increasing temperature the synthetic accessibility decreases as indicated by a higher SA score. Also, higher $T$s leads to an increase in the formation of radicals and a decrease in the number of heavy atoms in the generated molecules. The last two metrics correspond to premature termination of SELFIES parsing as invalid valencies would be produced, which truncates the generated molecules. As such, we observe that tuning the temperature $T$ during generation provides a way to control a trade off between sample diversity for synthetic accessibility. For the purposes of our GPT models, we consider $T = 1$ to produce reasonable molecules, though tuning $T$ may produce ideal sample sets for some desired downstream task.

\subsection{Transfer learning and inverse design of energetic materials}

The transfer learning procedure outlined in Section \ref{sec:training} enables X-GPT models that generate molecules with properties more characteristic of energetic materials than the corresponding base $\chi$hem-GPT model (Table \ref{table:x_gpt}). Crucially, fine-tuning preserves validity, demonstrating that the underlying model does not forget molecular grammar. Fine-tuning, particularly of GroupSELFIES models, lowers novelty as can perhaps be expected for training on a targeted molecular distribution. LoRA mitigates much of this novelty loss as fewer parameters are fine-tuned, presenting a means to get targeted generation while retaining distinction from the supervised training set. 

For unconditional generation, we observe that the mean detonation velocity and pressure shift toward higher values, e.g., from $3.32$ to $4.11$ km/s and from $3.95$ to $6.51$ GPa for the standard SELFIES $\chi$hem- to X-GPT models (Table ~\ref{table:x_gpt}). Here, we note that one should not expect a large overall distributional shift as the fine-tuning dataset contains molecules with modest expected performance. Instead, fine-tuning markedly improves the model’s ability to sample rare, high-performance candidates, as evidenced by the increased spread of detonation properties and a greater right-hand tail in the $(v, P)$ distributions, indicating more high-quality candidates above a given lower bound. We further observe a shift of molecular structure \textit{away} from the initial pretraining dataset and toward the supervised finetuning dataset as the KL scores for X-GPT SELFIES is $0.633$ to X-17K and $0.560$ to SAFE-8M while those for the GroupSELFIES variant are, respectively,  $0.802$ and $0.595$. Returning to the distinction between GroupSELFIES and SELFIES, we once again find greater quality in the GroupSELFIES models despite a modest decrease in validity. For instance, the GroupSELFIES $i$ encoding proves superior in producing molecules with both high detonation performance \textit{and} synthetic accessibility (Fig.~\ref{fig:sa_comparisons}b). While the peak of the synthetic accessibility to performance distributions shows minimal shift, the tail for high detonation velocity (Fig.~\ref{fig:sa_comparisons}) presents more generated examples for $X$-GPT compared to $\chi$hem-GPT.
\begin{table}[b!]
\subfloat[][]{
\begin{tabular}{llcccc}
Model & Encoding & Condition & Validity ($\uparrow$)  & SA Score ($\downarrow$)  & Novelty ($\uparrow$)  \\
\hline
$\chi$hem-GPT & SELFIES & None& $0.999\pm 0.001$  & $4.63\pm1.17$& $ \mathbf{0.998\pm0.001}$\\
 & GroupSELFIES $i$& None  & $0.911\pm0.006$ & $3.51\pm0.99$ & $0.995\pm 0.001$ \\
\hline
X-GPT & SELFIES &None&$\mathbf{1.000\pm 0.000}$  & $4.46\pm 1.10$ & $0.994\pm0.002$\\
(basic) & SELFIES& High $v$,$P$ & $\mathbf{1.000 \pm 0.001}$ & $4.35\pm1.13$ & $0.994\pm0.002$\\
 & GroupSELFIES $i$ & None& $0.931 \pm 0.006$ & $3.35\pm0.97$ & $0.973\pm0.004$\\
 & GroupSELFIES $i$ & High $v$,$P$ & $0.923\pm0.005$ & $\mathbf{3.34 \pm 0.98}$ & $0.975\pm 0.005$\\
\hline
X-GPT & SELFIES & None & $0.999\pm 0.001$ & $4.61 \pm 1.11$ &$0.997\pm0.003$\\
(LoRA) & SELFIES &High $v$,$P$ &$0.999\pm 0.001$ & $4.50 \pm 1.09$ & $0.995\pm0.003$\\
& GroupSELFIES $i$ & None &$0.929\pm0.012$ & $3.46\pm 0.99$ & $0.990\pm0.005$\\
& GroupSELFIES $i$ & High $v$,$P$ & $0.934\pm0.008$ &$3.42 \pm 0.99$&$0.990\pm0.002 $\\  
\end{tabular}
}\\
\subfloat[][]{
\begin{tabular}{llccc}
Model & Encoding & Condition & KJ $v$ (km/s) ($\uparrow$) & KJ $P$ (GPa) ($\uparrow$) \\
\hline
$\chi$hem-GPT & SELFIES & None& $3.318\pm 0.743$ & $3.950 \pm 1.973$\\
& GroupSELFIES $i$& None  & $3.274 \pm  0.775$ & $3.944 \pm 1.817$\\
\hline
X-GPT & SELFIES &None& $4.113 \pm 1.049$ & $6.505 \pm 3.672$\\
(basic) & SELFIES& High $v$,$P$& $\mathbf{4.350\pm1.077}$ & $\mathbf{7.426\pm4.463}$ \\
 & GroupSELFIES $i$ & None& $4.211 \pm 1.048$ & $6.968\pm4.081$\\
 & GroupSELFIES $i$ & High $v$,$P$ & $4.164\pm1.068$ & $6.840 \pm 3.957$\\
\hline
X-GPT & SELFIES & None &$4.109\pm0.934$ & $6.383 \pm 3.233$\\
(LoRA) & SELFIES &High $v$,$P$  & $4.315 \pm 1.124$ & $7.296\pm4.348$ \\
& GroupSELFIES $i$ & None &$4.087 \pm 0.969$ &$6.420\pm 3.476$\\
& GroupSELFIES $i$ & High $v$,$P$ & $4.145 \pm 1.070$&$6.744\pm3.929$\\  
\end{tabular}
}
\caption{Comparison of generated outputs between the base and fine-tuned GPT models. a) General metrics from all generated molecular texts across multiple machines ($N = 5000$). b) Kamlet-Jacobs \cite{kamlet1968chemistry} (KJ) metrics as characterized on a subset of the molecules in (a), filtered by converged calculations ($185\leq N \leq 530$). The condition for high $P$, $v$ corresponds to $(p,v) = (40 \ \text{GPa},10 \ \text{km/s})$. Standard fine-tuning is where the first and final transformer layers are unfrozen alongside the token output layer, while LoRA defines low-rank adaptation matrices with rank dimension $r= 8$ and $\alpha = 32$. Statistics follow the procedure outlined for Table \ref{tab:xhem_gpt}.}
\label{table:x_gpt}
\end{table}
\begin{figure}[b!]
\includegraphics[scale=1]{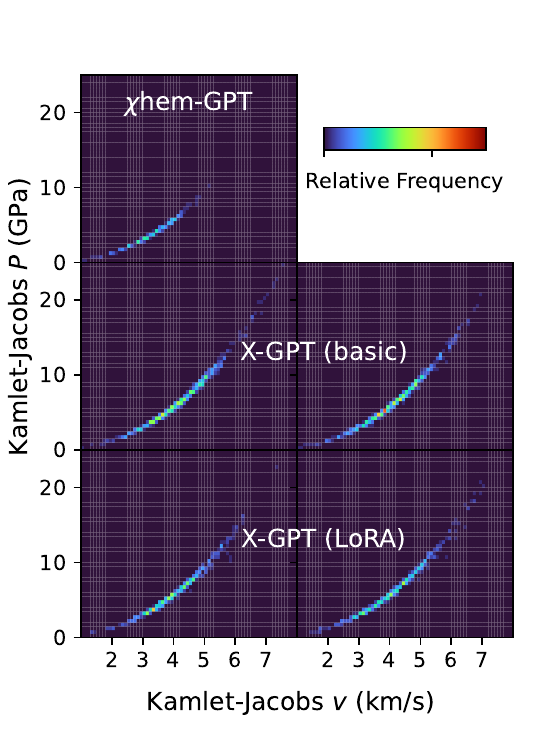}
\caption{Comparison of estimated detonation velocities and pressures from Kamlet-Jacobs equations \cite{kamlet1968chemistry} for unconditioned (left column) and conditioned generation (right) for fine-tuned GroupSELFIES models compared to the base model. All subfigures have been normalized to an identical maximum value.}
\label{fig:v_vs_p_comparisons}
\end{figure} 

As the X-GPT series was fine-tuned with input detonation $(v,P)$ property vectors, we obtain further conditioned generation by concatenating such property vectors onto it, further altering the output distributions. Such conditioned generation can shift the distributions of predicted velocities and pressures, albeit not always and not fully to that of the input property vector (see Table \ref{table:x_gpt}, Fig. \ref{fig:v_vs_p_comparisons}). The X-GPT models generate relatively more molecules at combined high $v$ and $P$, indicating better prospects for EM materials discovery. From the aggregate statistics, the SELFIES models exhibit a greater shift in distributions compared to the GroupSELFIES models (e.g., detonation property conditioning shifts mean Kamlet-Jacobs $v$ from $4.11$ to $4.35$ km/s for SELFIES but from $4.21$ to $4.16$ km/s for GroupSELFIES), which do not exhibit a similar positive shift (Table \ref{table:x_gpt}). Beyond aggregate statistics and analyzing the entire distribution of detonation properties, we note that the GroupSELFIES models display a modestly larger tail of high $v$ and $P$ for conditioned generation (Fig. \ref{fig:v_vs_p_comparisons}). For instance, out of $5000$ generated molecules, the basic GroupSELFIES X-GPT produced $348$ molecules with $v > 7 \text{km/s}$  as estimated by XChemProp when  X-GPT was conditioned with $(p,v) = (40 \ \text{GPa},10 \ \text{km/s})$, but only $224$ otherwise (the corresponding SELFIES X-GPT statistics are, respectively, $375$ and $226$). Despite producing relatively more high-performance EM candidates, this study also shows the limitations of fine-tuning on concatenated property data, which proved unable to dramatically shift generated distributions. These results for Kamlet-Jacobs benchmarking are broadly echoed by benchmarking on CHEETAH and XChemProp predicted detonation properties, albeit with distributional shifts (see Supplementary Materials Figs.~\ref{fig:v_vs_p_cheetah} and ~\ref{fig:v_vs_p_chemprop}).
\begin{figure}[b!]
\includegraphics[scale=1]{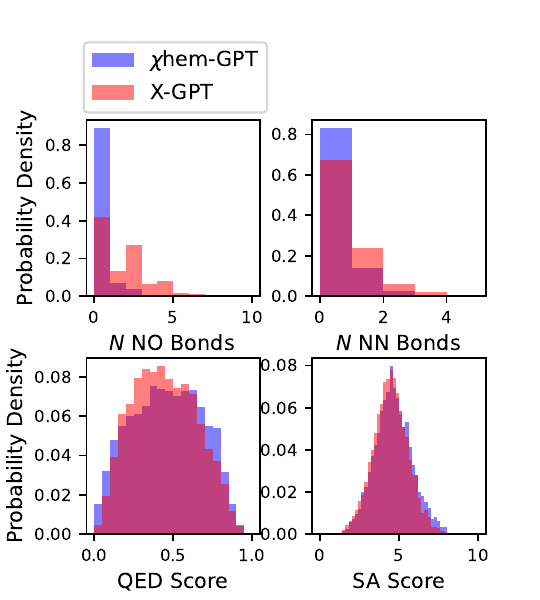}
\caption{Distributions of a) number of nitrogen-oxygen bonds, b) number of nitrogen-nitrogen bonds, c) quantitative estimation of druglikeness (QED), and d) synthetic accessibility score (SA Score) in large SELFIES-based GPT models.}
\label{fig:no_histogram}
\end{figure}

\begin{figure}
    \includegraphics[scale=  .6]{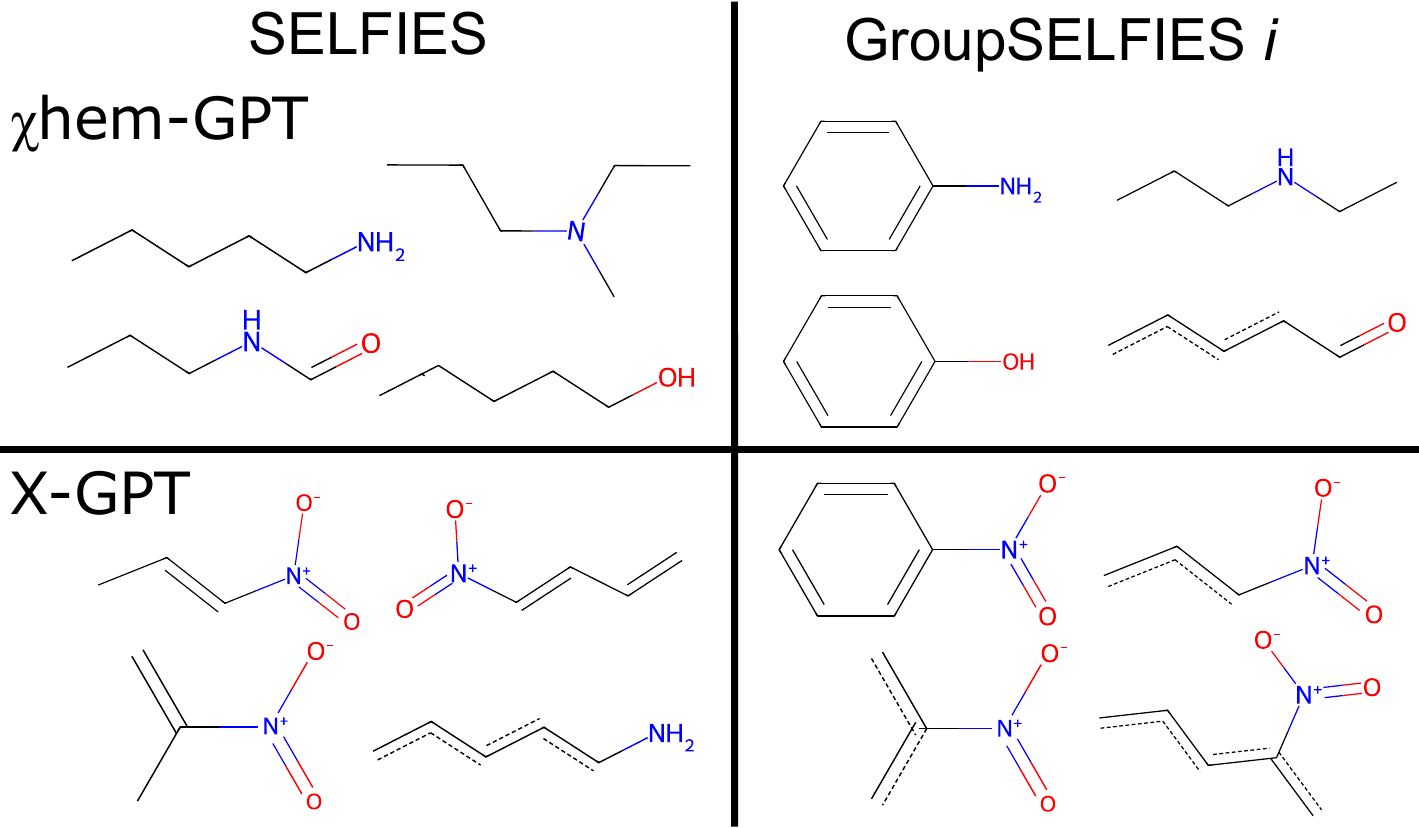}
    \caption{Common substructures of generated output from chemical language models that have been manually selected for diversity. Substructures were chosen as representative samples from the 200 most common subgraphs between size $5-10$ as obtained via RDKit \cite{rdkitsoftware}.}
\label{fig:substructures}
\end{figure}

Other metrics known to correlate with EM performance are observed in the X-GPT outputs, for instance, the number of N-O bonds, which are present in the nitro groups of many explosives, increases for X-GPT relative to $\chi$hem-GPT (Fig. ~\ref{fig:no_histogram}). Such a shift indicates that the models produce chemical features associated with energetic materials. Upon querying the generated molecules for common substructures, one observes that nitro groups are characteristic of the X-GPT outputs while they are not for the base foundational models (Fig. ~\ref{fig:substructures}). As with the aggregate detonation properties, one has to select from the tail of such properties to find N-O bond counts similar to known EMs (HMX, for instance, has eight N-O bonds). As transfer learning works from the entirety of the fine-tuning dataset, which includes molecules with low calculated expected performance (see Supplementary Materials Fig. ~\ref{fig:x_dist}), the tuned model necessarily includes such behavior in its outputs. Conditioning with property vectors ameliorates this, but we have not yet found an invocation of traditional supervised fine-tuning that fully permits the GPT model to consistently output high-performance outliers.

\section{Conclusion}

We have developed a class of GPT models for chemistry with a particular focus on energetic materials discovery. We have demonstrated how data oriented for pharmaceuticals can be used for unsupervised pretraining of neural network models that are later fine-tuned to energetic materials. Secondarily, we have shown the effectiveness of developing custom high-quality EM datasets incorporating high-throughput calculations for generation of specialized molecules. While fine-tuning can shift the underlying generated distribution in desirable directions, solely using supervised fine-tuning is limited by the underlying fine-tuning dataset. To mitigate this issue, one might use the best candidates from CLM models to run additional calculations and generate improved datasets, iteratively shifting the supervised training distribution. In contrast, reinforcement learning inspired approaches such as proximal policy optimization \cite{schulman2017proximal} and direct preference optimization \cite{rafailov2023direct} would use preferred examples of behavior in the training distribution to bias the CLM, effectively rewarding desired molecules and punishing undesired ones. Just as large language model alignment presents an ongoing challenge \cite{wang2024comprehensive}, so too does producing CLMs that generate novel desired chemical systems and not undesired ones.

In comparing the fragment GroupSELFIES encoding scheme to the more frequently utilized SELFIES, we observe a number of improvements in performance. We note that GroupSELFIES generated outputs with greater synthetic accessibility in contrast to SELFIES, though target properties such as detonation velocity are broadly equivalent following fine-tuning. Furthermore, GroupSELFIES offer a means to shift compute into CPU data processing prior to training and post-generation and away from GPU utilization during inference since fewer tokens must be predicted. With regards to the loss in generation validity, it appears that the decrease in validity is due to the model not distinguishing between structural and indexing tokens. As recently articulated \cite{skinnider2024invalid}, generating a modest amount of invalid molecules can serve as a simple pre-screening metric as schemes such as SELFIES that enforce validity produce improbable samples where they otherwise would have been invalid. Altering the index alphabet to one that does not fully overlap with the base lexicon for SELFIES presents a mechanism by which one can observe where a CLM fails to distinguish indexing from structural/elemental tokens. Such an encoding scheme could provide the regular grammar of SELFIES while avoiding certain undesirable skewing of the generated output distributions. 

Here, we have demonstrated the effectiveness of transfer learning for fine-tuning a CLM trained on a massive and dataset of generic small molecules to a narrower dataset of EMs. Via such post-training, X-GPT mimics a target EM distribution with some control over conditioned generation. While shifting molecular distributions can move generated molecules to a more desirable region of chemical space, further goals often entail that the generated molecules are uniquely performant along defined characteristics. That is, transfer learning on its own appears limited for an EM discovery task where one desires higher detonation pressures and velocities than the fine-tuning dataset, with only the tails of the distributions providing strong relative candidates. Reinforcement learning, which has so far been only recently applied to CLMs \cite{novo2024ye}, offers a method to shift the model outputs beyond their training dataset. A reward model that balances desirable EM properties, e.g., detonation performance and sensitivity, would be potentially able to guide fine-tuning into an even more desirable region of chemical space for EM discovery than X-GPT has reached. 

CLM development continues at a great pace, and here we have presented the adaptation and applicability of techniques developed predominantly for drug discovery toward energetic materials discovery. Pretrained neural networks enable the reuse of computational expenditure in transferring to many potential specific domains, though CLMs have been primarily applied to pharmaceuticals. Our demonstration of X-GPT extends generative methods toward energetic materials, bolstering an expanding body of literature in supervised fine-tuning of chemical language models for targeted applications ranging from catalyst discovery \cite{mok2024generative} to battery electrolyte design \cite{soares2023capturing}.
\section{List of Abbreviations}
\begin{description}[leftmargin=*, widest=DCCHTM]
\item[CLM]
Chemical Language Model
\item[CPU]
central processing unit
\item[EM]
energetic material
\item[FOX-7]
1,1-diamino-2,2,-dinitro-ethylene
\item[GPT]
generative-pretrained transformer
\item[GPU]
graphics processing unit
\item[KL]
Kullback-Leibler (divergence/score)
\item[LLM]
Large Language Model 
\item[LoRA]
Low-Rank Adaptation
\item[ML]
machine learning
\item[RDX]
cyclotrimethylene trintiroamine (hexogen)
\item[SELFIES]
SELF-referencIng Embedded Strings
\item[SMILES]
Simplified Molecular Input Line Entry System
\item[TNAZ]
1,3,3-trinitroaze-tidine
\item[TNT]
trinitrotoluene
\end{description}
\section{Declarations}

%
\subsection{Competing interests}
The authors declare no competing or conflicts of interest for this study.

\subsection{Author contributions}
W.K.K, I.M, and C.S. conceived the project. A.S. developed and trained the GPT models with inputs from C.G.C. and W.K.K., as well as analyzed model benchmarking data. R.S.U., M.C.D., and I.M performed density functional theory and detonation performance calculations of the X-17K dataset. A.S. and W.K.K. developed the chemical property surrogate models. A.S. wrote the manuscript with inputs from C.G.C., I.M., M.J.C., and W.K.K. All authors revised and approved the final version of the paper. I.M. and W.K.K. supervised the overall development of the computational framework.

\subsection{Funding}
Research presented in this article was supported by the Laboratory Directed Research and Development program of Los Alamos National Laboratory under project number LDRD 20250006DR. This research used resources provided by the Los Alamos National Laboratory Institutional Computing Program, which is supported by the U.S. Department of Energy National Nuclear Security Administration under Contract No. 89233218CNA000001.


\newpage

\clearpage
\setcounter{section}{0}
\setcounter{figure}{0}
\setcounter{table}{0}
\makeatletter
\renewcommand \thesection{S\@arabic\c@section}
\renewcommand \thefigure{S\@arabic\c@figure}
\renewcommand \thetable{S\@arabic\c@table}
\makeatother

\begin{center}
\large{\textbf{Supplementary Materials for Generative Chemical Language Models for Energetic Materials Discovery}}
\end{center}

\tableofcontents
\section{Notes on Software}
During development of this paper, the Python software for results underwent numerous changes and updates to the Python environment. Final benchmarking runs were performed in a largely stable environment, with the version numbers provided in the main text and here being for a representative benchmarking run.

Where possible for benchmarking, we used open-source packages to ensure availability of those methods to interested readers. That noted, we have made some updates to the used packages, summarized here. Namely, as noted in the main text, GroupSELFIES \cite{cheng2023group} has been forced to use the SELFIES \cite{krenn2020self} indexing alphabet in the default case. Furthermore, GroupSELFIES have had their serialization improved so as to be pickle serializable. With regards to Kullback-Leibler divergence \cite{kullback1951information}, we refer to the parameters and descriptors for Guacamol \cite{brown2019guacamol}, though we did update the handling of discrete histograms to permit updated dependencies. ChemProp \cite{heid2023chemprop} has been minimally altered, though its data processing has also been updated to be compatible with models trained using a prior version of it.

\subsection{PyTorch}
All GPT models were developed using the PyTorch \cite{paszke2019pytorch,pytorchsoftware} framework, designed to work on GPUs and on a x86-64 Linux architecture. Distributed data parallel training was performed via PyTorch Lightning\cite{pytorchlightningsoftware} and tracked via Aim \cite{aimsoftware}. This SI is not the definitive source for versioning information, though the cited versions should be broadly correct. 

\section{Dataset Characterization}
\begin{table}
\begin{tabular}{l|c|c|c}
 Name    & SELFIES Size &GroupSELFIES Size & Sources \\
SAFE-8M  & $7897563$ & $7905990$ &\cite{noutahi2024gotta,irwin2005zinc,chambers2013unichem} \\
X-17K & $17407$& $17389$ &\cite{allen2002cambridge}\\
\end{tabular}
\caption{Overview of datasets used in this study. Note that for the dataset size, this is of the entire dataset prior to train/val/test splitting, and final batches smaller than the batch size are dropped during training. Note that datasets differ trivially in exact size and composition due to rare parsing issues in constructing the datasets from initial SMILES.}
\label{tab:si_datasets}
\end{table}
Below are plots characterizing the datasets our chemical GPTs have trained on. 

\subsection{SAFE-8M}
\label{si:sec_safe_dataset}
SAFE-8M may be accessed from \href{https://huggingface.co/datasets/datamol-io/safe-gpt}{Hugging Face}, where it is processed by converting the data in \mono{test-00000-of-00013-a7e083cf8af49982.parquet} into SMILES and then into SELFIES variants.\\
\\
For SAFE-8M, we have sampled $1\%$ of the underlying dataset for further characterization of molecular features. Some outliers may be omitted from the histograms. 
\begin{figure}
\includegraphics[scale=0.7]{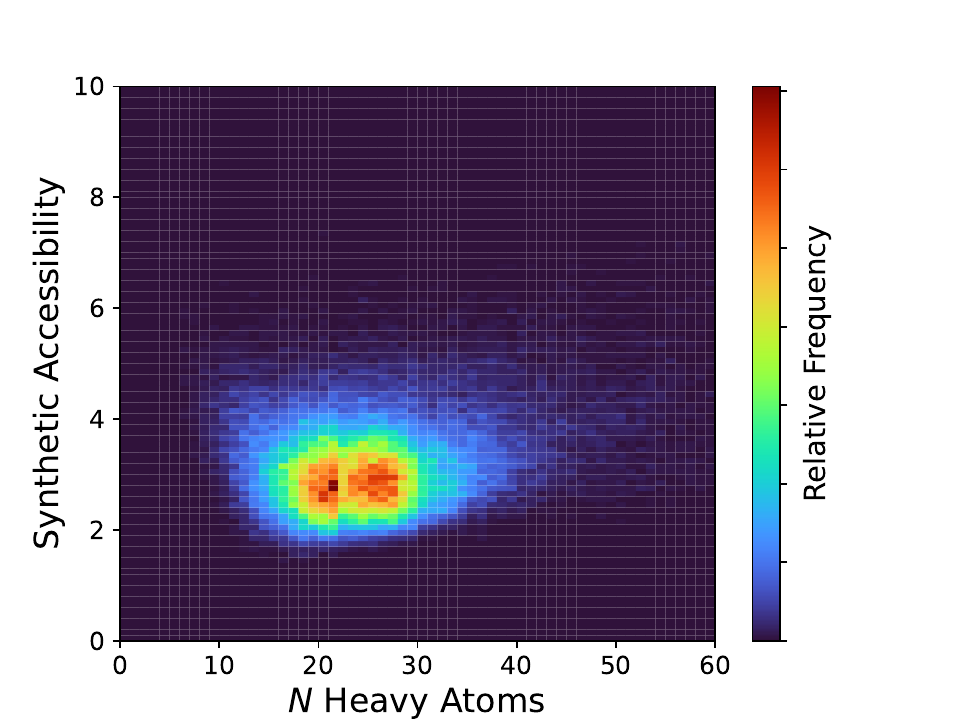}
\caption{Distribution of synthetic accessibility to heavy atoms in molecule for SAFE-8M.} 
\label{fig:safe_dist}
\end{figure}
\begin{figure}
\includegraphics[scale=0.7]{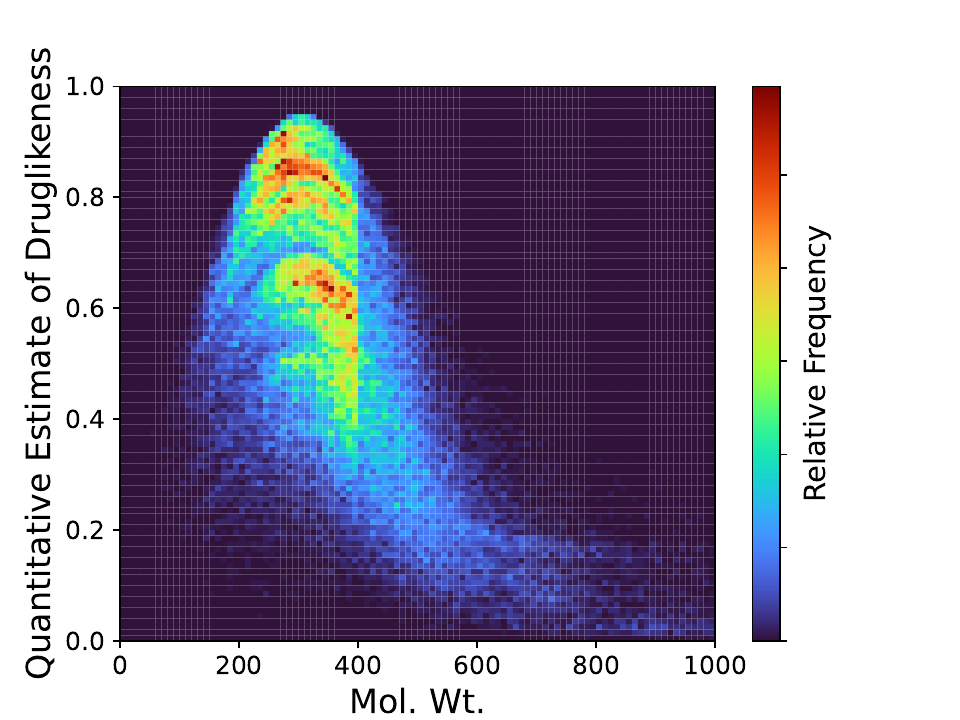}
\caption{Distribution of druglikeness to molecular weight in molecule for SAFE-8M.} 
\label{fig:safe_dist_qed}
\end{figure}

\newpage
\subsection{X-17K}

For X-17K, we have characterized the entire dataset. Some outliers may be omitted from the histograms.
\begin{figure}[h!]
\includegraphics[scale=0.7]{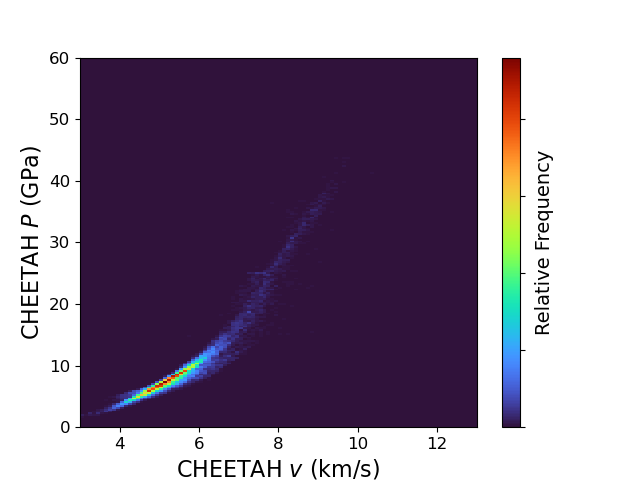}
\caption{Calculated velocities and pressures from CHEETAH\cite{fried1994cheetah} for X-17K.}
\label{fig:x_dist}
\end{figure}
\section{Model Parameters}
\label{sec:si_model_params}
\begin{table}
\begin{tabular}{l|c|c|c|c}
Model & Small (SELFIES) & Small (GSELFIES) & Large (SELFIES) & Large (GSELFIES) \\
\hline
Parameters & $40$ M& $40$ M & $150$ M & $160$ M\\
$N$ Layers & $12$ & $12$ & $12$ & $12$ \\
$d_{embed}$ & $512$ & $512$ & $1028$ & $1028$ \\
$w_{context}$ & $256$ & $256$ & $512$ & $512$\\ 
$n_{head}$ & $4$ & $4$ & $4$ &$4$  \\ 
vocab. size & $975$ & $1617$ &  $975$& $1617$\\
$p_{dropout}$ & $0.01$ & $0.01$ &  $0.01$ & $0.01$
\end{tabular}
\caption{Model architecture parameters. Parameters are rounded to the nearest $10$ million and obtained via a summation of all registered parameters and vocabulary size includes special tokens.}
\label{tab:si_model_parameter}
\end{table}
\section{Training}
\label{sec:si_training}
\begin{figure}
\includegraphics[scale=0.7]{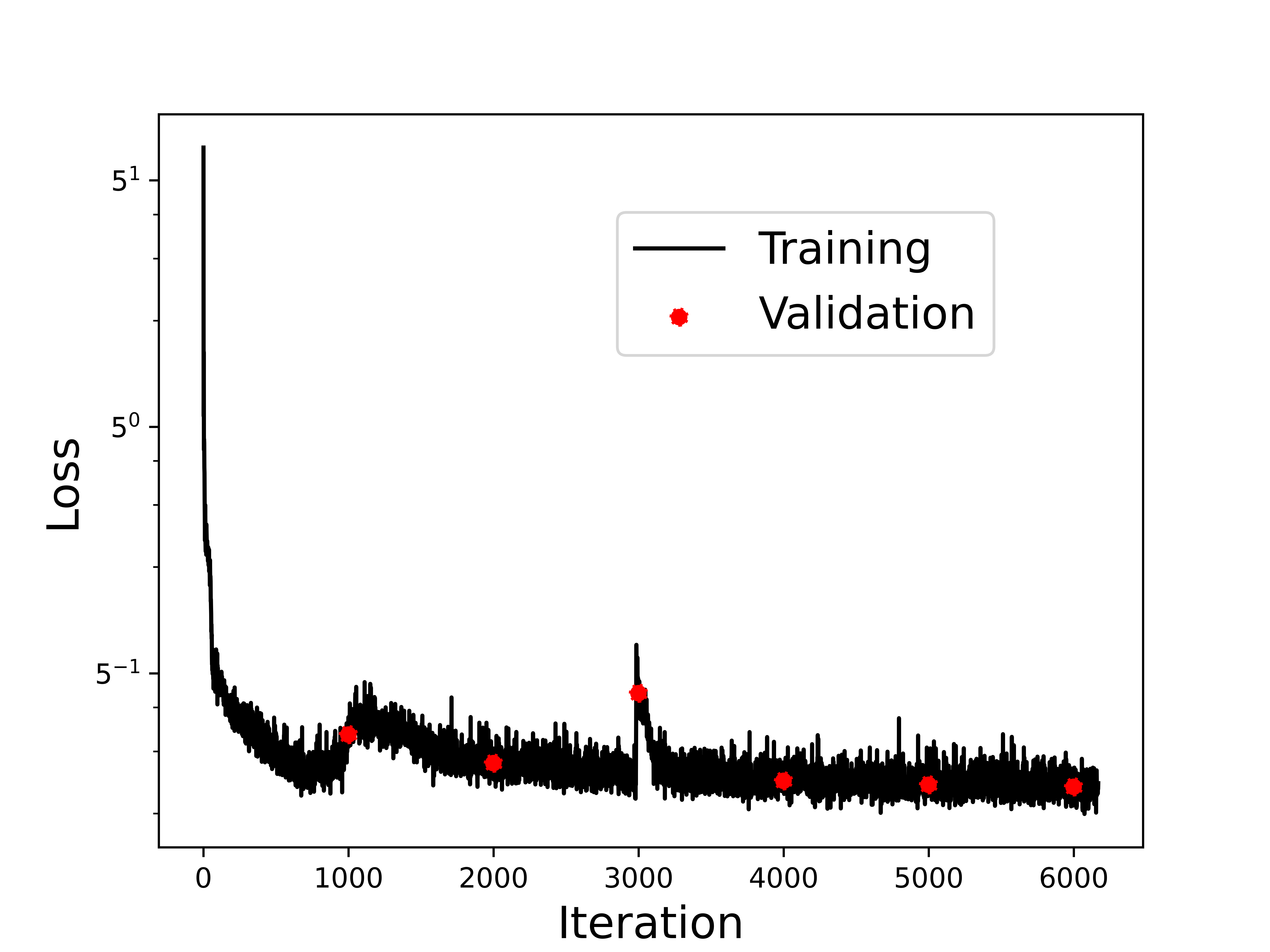}
\caption{Training and validation loss for $\chi$hem-GPT (LARGE, SELFIES).}
\end{figure}
\begin{figure}
\includegraphics[scale=0.7]{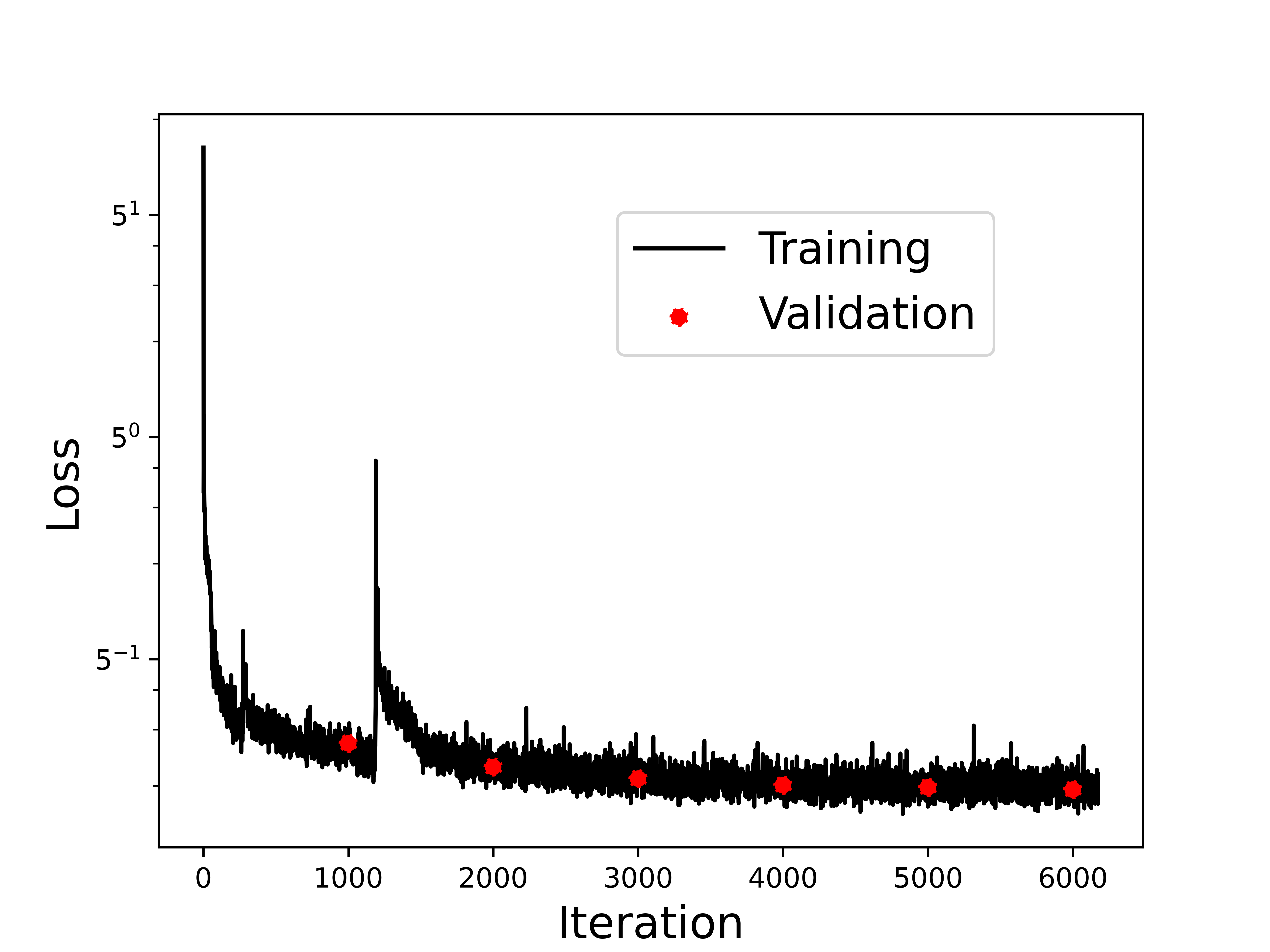}
\caption{Training and validation loss for $\chi$hem-GPT (LARGE, GroupSELFIES).}
\end{figure}
\begin{figure}
\includegraphics[scale=0.7]{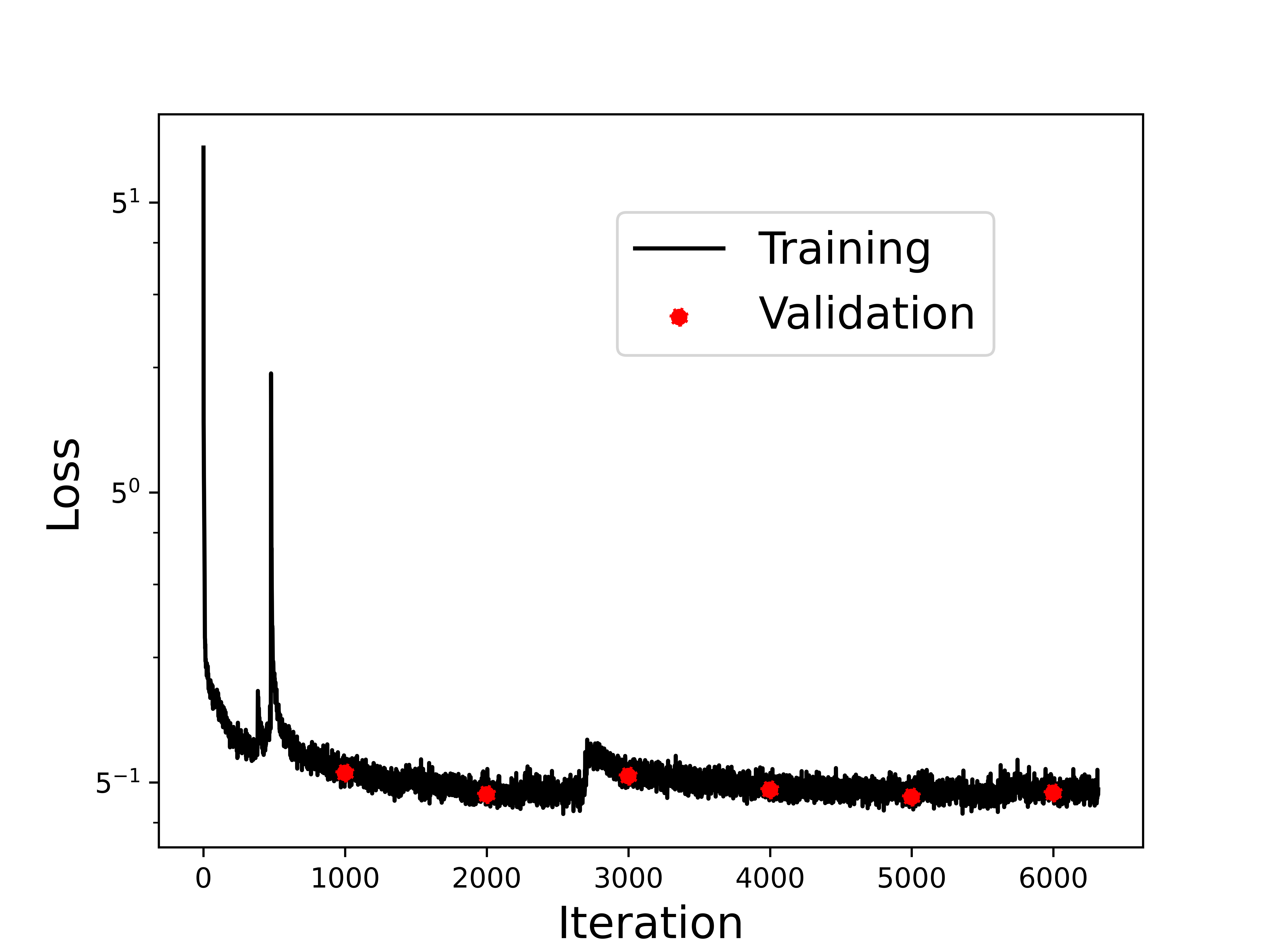}
\caption{Training and validation loss for $\chi$hem-GPT (SMALL, SELFIES).}
\end{figure}
\begin{figure}
\includegraphics[scale=0.7]{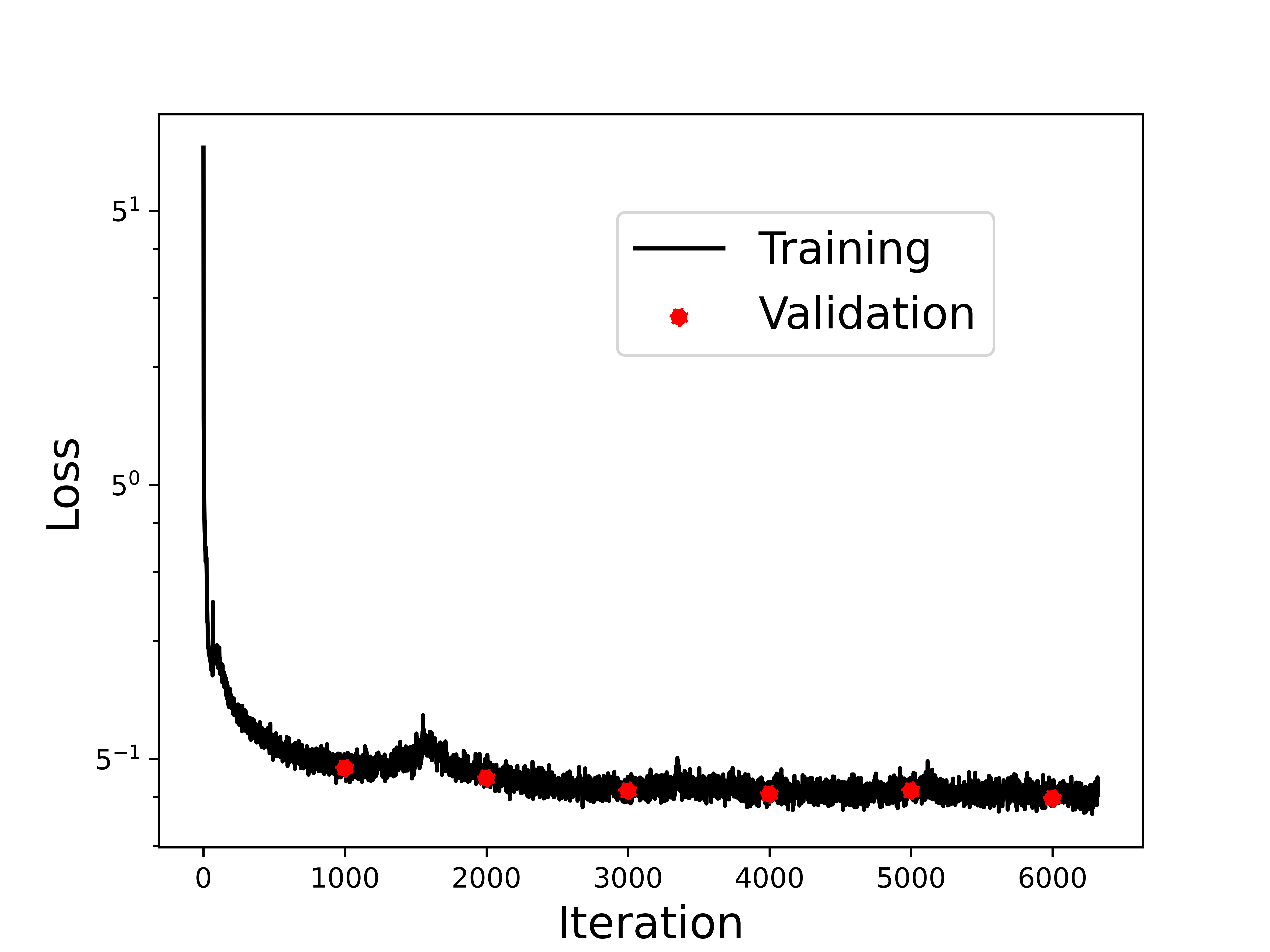}
\caption{Training and validation loss for $\chi$hem-GPT (SMALL, GroupSELFIES).}
\end{figure}

\newpage
\section{Surrogate ChemProp Model}
\begin{figure}[t!]
\includegraphics[scale=0.8]{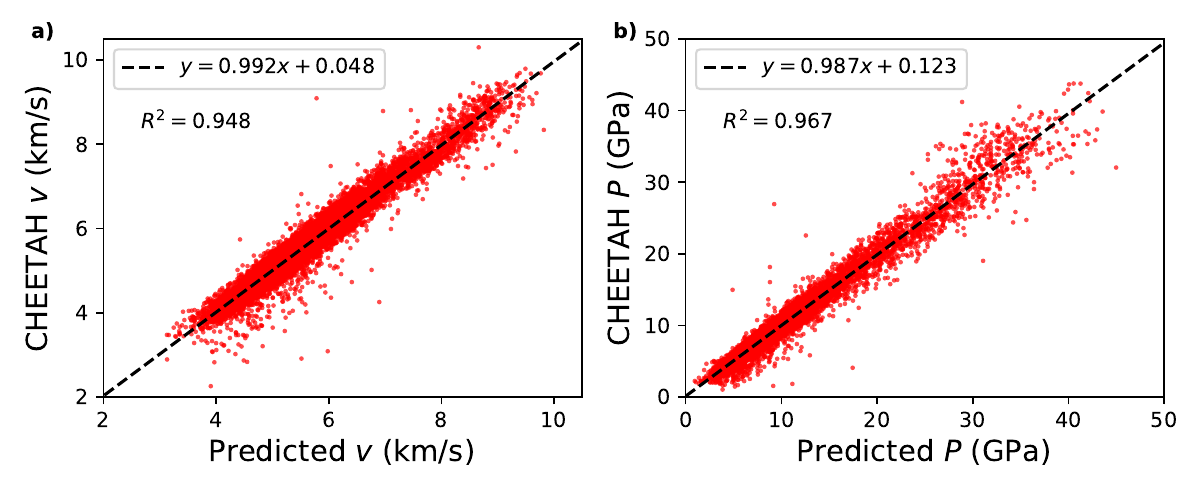}
\caption{Validation of ChemProp model for detonation a) velocities and b) pressures for the X-17K dataset.}
\label{fig:xchemprop}
\end{figure}

\begin{figure}[h!]
\includegraphics[scale=.9]{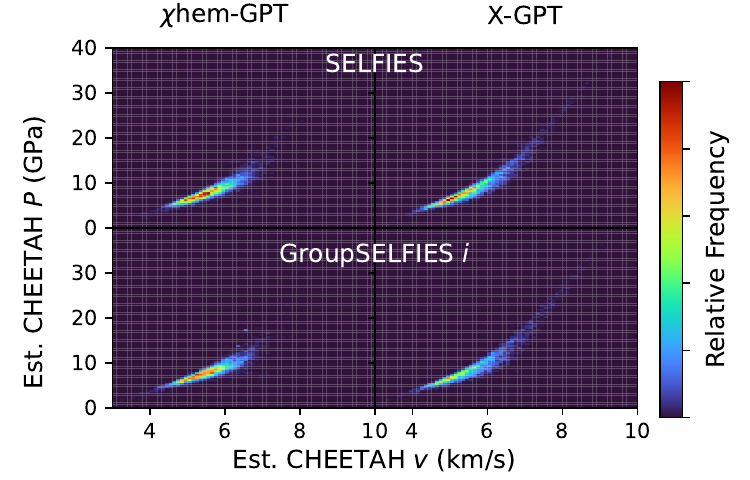}
\caption{Comparison of estimated detonation velocities and pressures via XChemProp for unconditioned generation of $\chi$hem- and X-GPT (standard fine-tuning) models. All subfigures normalized to an identical maximum value.}
\label{fig:v_vs_p_chemprop}
\end{figure}

\begin{figure}[h!]
\includegraphics[scale = .9]{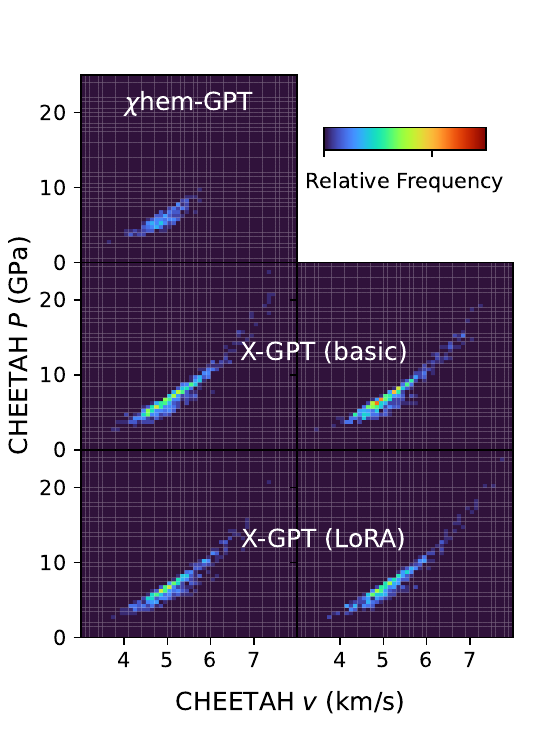}
\caption{Comparison of CHEETAH calculated (not estimated) detonation properties for  $\chi$hem- and X-GPT GroupSELFIES $i$ models. All subfigures normalized to an identical maximum value.}
\label{fig:v_vs_p_cheetah_group}
\end{figure}

\begin{figure}[h!]
\includegraphics[scale = .9]{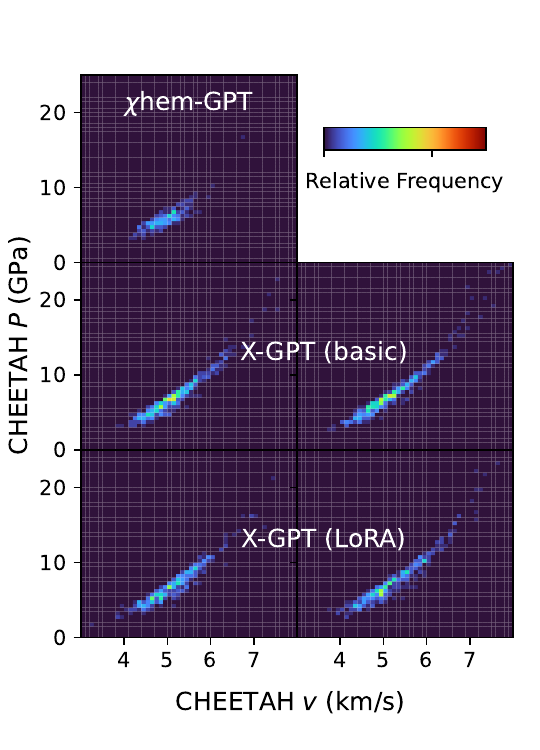}
\caption{Comparison of CHEETAH calculated (not estimated) detonation properties for  $\chi$hem- and X-GPT SELFIES models. All subfigures normalized to an identical maximum value.}
\label{fig:v_vs_p_cheetah}
\end{figure}

\begin{figure}[h!]
\includegraphics[scale = .9]{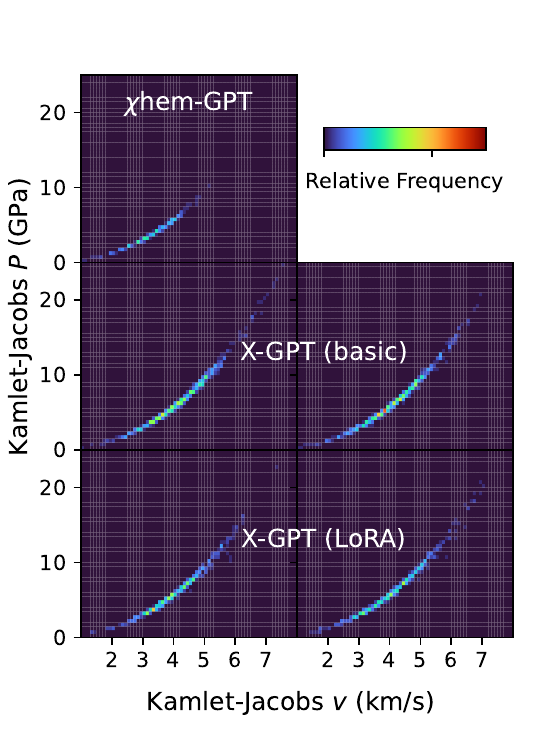}
\caption{Comparison of Kamlet-Jacobs calculated (not estimated) detonation properties for  $\chi$hem- and X-GPT SELFIES models. All subfigures normalized to an identical maximum value.}
\label{fig:v_vs_p_kj}
\end{figure}

As noted in the main text, we developed a surrogate ChemProp \cite{heid2023chemprop} model trained on CHEETAH \cite{fried1994cheetah} results for validation as part of our training development loop. We observe that it performs well on its training data (Fig. ~\ref{fig:xchemprop}), though it tends to predict larger detonation velocities (Fig. ~\ref{fig:v_vs_p_chemprop}) of generated samples compared to the density-functional theory validated subset characterized via Kamlet-Jacobs (Fig. ~\ref{fig:v_vs_p_kj}), a characteristic it shares with the CHEETAH validation data (Figs. ~\ref{fig:v_vs_p_cheetah}, ~\ref{fig:v_vs_p_cheetah_group}) to which the observed distributions are more similar. 
\section{Model Generation Characterization}
\label{sec:si_characterization}
We have included additional heatmaps depicting the generated distributions from the chemistry GPT models reported in the main text.
\begin{figure}
 \includegraphics[]{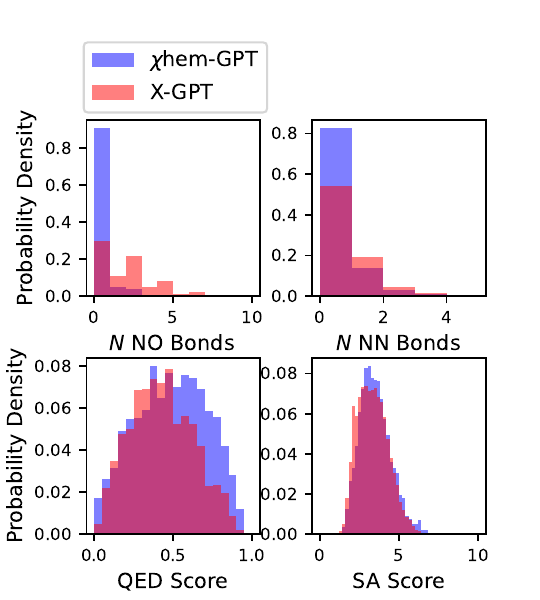}
 \caption{Distributions of a) number of nitrogen-oxygen bonds, b) number of nitrogen-nitrogen bonds, c) quantitative estimation of druglikeness (QED), and d) synthetic accessibility score (SA Score) in large GroupSELFIES $i$-based GPT models.}
 \end{figure}

\clearpage
\newpage
\bibliography{x_gpt.bib}

\end{document}